\documentclass[pdflatex,sn-mathphys-num]{sn-jnl}% Math and Physical Sciences Numbered Reference Style
\usepackage{graphicx}%
\usepackage{multirow}%
\usepackage{amsmath,amssymb,amsfonts}%
\usepackage{amsthm}%
\usepackage{mathrsfs}%
\usepackage[title]{appendix}%
\usepackage{xcolor}%
\usepackage{textcomp}%
\usepackage{manyfoot}%
\usepackage{booktabs}%
\usepackage{algorithm}%
\usepackage{algorithmicx}%
\usepackage{algpseudocode}%
\usepackage{listings}%
\usepackage{graphicx}
\usepackage{textcomp}
\usepackage{xcolor}
\usepackage{braket}
\usepackage{float}
\usepackage[normalem]{ulem}
\usepackage{booktabs}
\usepackage{placeins}
\theoremstyle{thmstyletwo}%

\theoremstyle{thmstylethree}%

\begin{document}
\title[Quantum-Assisted Optimization Guided by Machine-Learned Risk Maps]{Quantum-Assisted Optimization Guided by Machine-Learned Risk Maps for Aerial Surveillance Routing}
% \title{Quantum-Assisted Optimization Guided by Machine-Learned Risk Maps for Aerial Surveillance Routing}


\author*[1]{\fnm{Dorian} \sur{Lauwerier}}\email{dorian.lauwerier@usherbrooke.ca}
\author[2]{\fnm{Alexis} \sur{Vieloszynski}}
\author[1]{\fnm{Yves} \sur{Bérubé-Lauzière}}
% \email{iiiauthor@gmail.com}
\author[1]{\fnm{Victor} \sur{Drouin-Touchette}}

\affil[1]{\orgdiv{Département de génie électrique et de génie informatique}, \orgname{Université de Sherbrooke}, \orgaddress{\city{Sherbrooke}, \state{Québec}, \country{Canada}}}
\affil[2]{\orgdiv{Département d'informatique}, \orgname{Université de Sherbrooke}, \orgaddress{\city{Sherbrooke}, \state{Québec}, \country{Canada}}}
% \affil[3]{\orgdiv{Département de physique}, \orgname{Université de Sherbrooke}, \orgaddress{\city{Sherbrooke}, \state{Québec}, \country{Canada}}}

% \affil[2]{\orgdiv{Department}, \orgname{Organization}, \orgaddress{\street{Street}, \city{City}, \postcode{10587}, \state{State}, \country{Country}}}

% \affil[3]{\orgdiv{Department}, \orgname{Organization}, \orgaddress{\street{Street}, \city{City}, \postcode{610101}, \state{State}, \country{Country}}}


\abstract{Analog quantum computing with neutral atom computers is rapidly evolving into a promising way to perform combinatorial optimization tasks at scale. As technology matures, exploring real use cases leveraging such unique information processors becomes warranted; one example is logistics management where vehicles are assigned routes to perform a given task. Autonomous aerial vehicles (drones) are increasingly used for surveillance over a given territory. In this paper, we propose an innovative framework integrating classical machine learning forecasts and quantum annealing to find the best combination of tours for a given surveillance criterion. The risk maps produced by machine learning are used to find possible surveillance tours, while a quantum-assisted heuristic chooses the best combination, producing a full surveillance plan. By modelling our decision problem for drone paths as a Team Orienteering Problem (TOP) with additional constraints, we construct a new heuristic using quantum annealing to find good solutions. We apply this framework to wildfire prevention, where finding tours most likely to detect a starting fire has clear benefits. We study the effect of larger quantum resources on heuristic performance and compare those results to \textcolor{black}{the use of classical resources.}}


%%\pacs[JEL Classification]{90C59, 90C11, 90B06, 68Q12, 81P68}


\maketitle
\section{Introduction}\label{sec1}
Aerial surveillance is a task performed in various contexts: preventing oil spills, monitoring wild animal populations, enforcing environmental laws by detecting illegal activities, preventing wildfire outbreaks, or monitoring ongoing wildfires. 
With the recent technological development of fixed-wing Unmanned Aerial Vehicles (UAVs) and their decreasing cost, new applications in agriculture, mineral deposit prospecting, and archaeology could also emerge. 
While deciding on the optimal tour of a single aerial vehicle is an interesting problem by itself, a more realistic setting  has either vehicles making multiple tours per day or a fleet of vehicles each performing different tours.
Furthermore, if what we seek to observe does not depend on the start time within a given period (e.g., a day), making several tours one after another is equivalent to performing multiple tours simultaneously, thus both situations can be described by the same mathematical model. 

The difficulty of choosing a good set of tours for a fleet of aerial vehicles is often threefold. 
First, the area that an aerial vehicle has to observe is often vast and remote - deciding \textit{where} to look is a hard question in itself. 
Second, the data available for these remote areas is often insufficient to make accurate predictions: the event may be highly localized or infrequent while the studied area is very large, thus complicating the specification of \textit{what} to look for.
Third, the number of operations required to find the best combination of tours increases exponentially with tour length, making it extremely difficult to identify an optimal combination. 
Consequently, \textit{how} to solve this third problem optimally is perhaps the hardest question of all. 

In this paper, we propose a novel framework integrating classical machine learning forecasting and quantum optimization techniques to provide heuristic solutions by seeking the most effective surveillance routes for prevention tasks. 
Our methodology operates through a two-phase process. 
It first employs machine learning algorithms trained on historical data to generate accurate forecasts which must be expressed in the form of a map. 
Then, in the second phase, we model the selection of drone surveillance routes as an integer linear problem. 
Leveraging its structure and constraints, we design a heuristic that reduces our task to solving a Maximum Independent Set (MIS) problem on a conflict graph. 
Independent set problems such as MIS can be naturally mapped onto the Ising Hamiltonian of analog neutral-atom quantum processors, with demonstrated practical speed-ups for quantum computers over classical methods on hard instances~\cite{Ebadi2022MIS, Pichler2018MIS, wurtz2024industryapplicationsneutralatomquantum}. 
This property, along with the diversity of solutions generated by quantum computing approaches due to the inherent statistical nature of sampling from quantum states \cite{da2023quantum, PhysRevA.110.012434, perron2025leveraging}, motivates our use of analog quantum computing, and in particular quantum annealing, in this paper. 

Finally, we apply this framework to the routing of drones for wildfire prevention. 
Wildfires contribute significantly to global carbon emissions, accelerate biodiversity loss, and often result in extensive property damage and loss of life. 
This issue is strongly aggravated by climate change, creating global challenges. 
Effective wildfire prevention strategies are essential to minimize these impacts. 
Among the emerging technologies used to achieve this goal, Unmanned Aerial Vehicles (UAVs) equipped with thermal cameras~\cite{s23146635} are promising tools for early detection. 
However, monitoring vast, remote forested areas remains a considerable challenge, requiring either more drones or UAVs with longer range. 
The latter can be achieved with fixed-wing aircrafts, which have become increasingly affordable, thus allowing governmental organizations to adopt them for surveillance tasks. 
Therefore, the development of advanced decision-support systems for wildfire mitigation is becoming a growing priority for firefighting agencies. 

Our framework addresses both predictive accuracy and operational efficiency, demonstrating a predict-then-optimize pipeline adaptable to real-world operational contexts. 
By validating against a classical optimization method, we confirm comparable or improved performance in a short time, underscoring the practicality and value of quantum-assisted approaches in wildfire prevention logistics and other aerial surveillance tasks. 
This practical use case for quantum computing therefore offers a decision-support approach that could assist firefighting departments in improving wildfire surveillance planning.

\section{Related Works}\label{sec2}

\begin{figure}[h]  % Forces the figure to appear "Here"
  \centering
  \includegraphics[width=0.55\columnwidth]{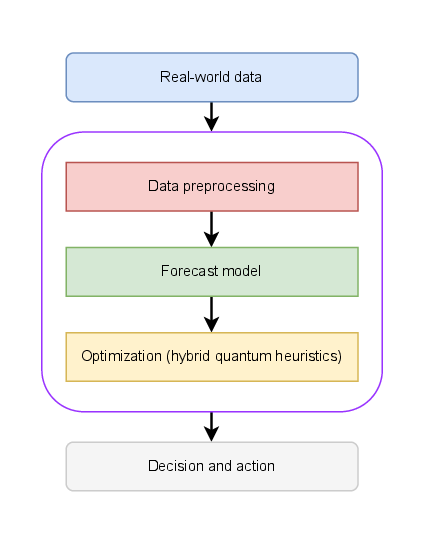}
  \caption{Our framework (purple) insert into a complete aerial surveillance workflow and help the decision making task of routing the drone fleet based on the predicted "reward" map.}
  \label{fig:1}
\end{figure}

The framework (see Fig.~\ref{fig:1}) we propose unifies terrain prediction and the subsequent decision regarding the maximization of a given observation goal. It requires a machine learning model for the forecasting process and an optimizer. 
This provides a complete pipeline from data collection to a final decision. 
It can be applied in situations where the parameters of the optimization problem have to be estimated. Nevertheless, the problem is not a stochastic optimization problem; once the parameters are fixed, the decision is no longer influenced by random fluctuations.

As drones are a recent technology, drone routing is itself a recent topic. UAV's detection abilities, cost, automony and range have improved a lot \cite{nguyen2022stateaerialsurveillancesurvey} in recent years, and they have been proposed for monitoring in remote areas. 
The modelling of such multi-drone routing can differ from a Vehicle Routing Problem (VRP) to a Team Orienteering Problem (TOP), with their respective variants \cite{bai2025dynamicunmannedaerialvehicle}.  
However, for most problems, finding the best paths for the fleet is not an easy task and presents many challenges in itself \cite{drones7030169}. Although we did not find an exact match for our specific drone path-planning problem in the literature, similar approaches model the environment as a uniform grid and often use fixed-wing drones due to their superior battery efficiency.

Decision-making and quantum computing have promising intersections; the example of quantum annealing for optimization is well studied for a variety of problems, in particular for the Maximum Independent Set (MIS). This problem can be naturally mapped to the combined effect of a laser to coherently excite neutral atoms to their Rydberg levels and the interactions between these atoms. Such a platform, called a  Neutral Atom Quantum Computer (NAQC), can be used for analog quantum computing where the laser's intensity and frequency is slowly changed so as to approximate the ground state of the system, which itself encodes an approximate solution to the MIS problem. The underlying motivation for such quantum methods are expected to be faster on hard instances than classical methods \cite{Ebadi2022MIS,Pichler2018MIS}, where a polynomial improvement on simulated annealing was observed. It has been demonstrated that NAQCs can also tackle other problems~\cite{wurtz2024industryapplicationsneutralatomquantum} either by reducing them to MIS or by using MIS as a component in a more global resolution method, such as column generation \cite{PhysRevA.110.012434, perron2025leveraging}. 
We note that a recent contribution proposes to solve drone path planning by combining quantum annealing and gate-based paradigms in a two-step approach \cite{osaba2025solvingdroneroutingproblems}. Our approach provides a distinct heuristic for drone path planning by instead combining classical and quantum resources.

This heuristic alone is not sufficient to provide good-quality decisions for aerial surveillance planning. Indeed, to make a decision, one needs predictions about which areas are interesting to observe (these are the initial parameters of the method to optimize over). These are not easy to obtain in general, so a forecasting method is needed, resulting in a map of the sought-after quantity (e.g. hazard, risk, vulnerability) required to correctly plan the aerial vehicle paths. 
In the example we delve into in this paper, this map represents forest fire ignition likelihoods~\cite{DeLaFuente2024IntegratedFireMonitoring}. 
Machine learning (ML) is the main technique used to predict fire risks and improve response strategies. By analyzing environmental data such as weather, vegetation, and past fires, ML models such as gradient boosting \cite{koh2023gradient} can provide accurate predictions. These models can also estimate the likelihood of fire ignition for specific locations, allowing authorities to focus prevention efforts on high-risk areas. This helps with early warning systems and resource planning, enabling faster responses and reducing the impact of wildfires. Studies like Refs.~\cite{koh2023gradient,xie2019forest} showcase the value of ML in enhancing fire management. We note that the XGBoost model, which we use, has already been successfully used in wildfire forecasting in Portugal \cite{hu2025xgboostmeetsinlatwostage}.

Finally, several frameworks have been developed to assess wildfire risks \cite{scott2013wildfire}, to prevent and mitigate them \cite{nhess-16-643-2016}, and to manage wildfires \cite{ALCASENA2019303}. All of these frameworks imply not only technical knowledge and techniques but also politics, population information, and socioeconomic considerations. This falls under the Integrated Fire Management (IFM) process \cite{OliverasMenor2025}. The IFM strongly encourages planning methods and risk assessment, as both are necessary to efficiently prevent wildfires and reduce their impacts. However, while risk assessment is widely discussed in the literature, planning and logistics around such risks remain less explored despite their critical importance. 

\section{Model Framework}\label{sec3}

\subsection{Forecasting with Machine Learning}
We employ machine learning (specifically, XGBoost, though other methods could be used) to predict risks or rewards based on spatially distributed environmental data. Such data can be acquired in various ways: satellite, weather stations, hygrometers, etc. It is then preprocessed to be amenable to analysis. The study area is divided into a grid, where each cell represents a specific geographical location. For each cell, the ML model is trained on various meteorological factors and learns patterns that influence the local forecast objective value.

In the case of wildfire prediction, this local value is the probability of a wildfire occurring in a given cell on the following day. By leveraging historical data and recognizing patterns in environmental conditions, the model can therefore anticipate high-risk areas and support proactive decision-making for fire prevention and resource allocation.

\subsection{An optimization model for drone surveillance}\label{optModel}

We model the optimization sub-problem as a single drone which can make $K$ tours during a day. The goal is for the tours to visit locations on the forecast map such that the sum of probabilities along the path to be maximized.
A trivial solution to this formulation is to repeatedly return to the area where the event is most likely to happen. However, other areas with similar (but lower) probabilities could also be interesting to observe. As a consequence, we add in our model a constraint of maximal overlap between tours such that the candidate solutions are also spatially diverse. Note that this spatial objective could also be added in a bi-objective problem. However, such problems are notoriously harder to tackle.
% and there is no  quantum computers are efficient in this task.  
The problem then belongs to the class of Orienteering Problems (OP) \cite{https://doi.org/10.1002/1520-6750(198706)34:3<307::AID-NAV3220340302>3.0.CO;2-D}, in particular it is a Team Orienteering Problem (TOP) \cite{CHAO1996464} with an additional overlap relaxation. Here, the "teams" are the number $K$ of tours that a drone can make daily. 

Mathematically, we model the problem as a graph $G = (V,E)$, with $V$ a set of vertices representing the possible areas the drones can visit (we choose vertex $1$ as the start and end of the tours, i.e. the drone depot), $E$ a set of edges between vertices connecting neighbouring areas on the grid. $n = |V|$ is the number of possible locations. We have the following parameters: $\phi_i$ is the probability of observed at node $i$, $D_{ij}$ is the distance between vertices $i$ and $j$, $D_{\max}$ is the maximal distance a drone can fly during a tour and finally $T$ is the overlap threshold, i.e. the number of common vertices that any two tours can have. With the following decision variables,
 
\begin{alignat*}{2}
% &\textbf{Decision variables:} \\
&\psi_{ij}^k \in \{0,1\} && \text{1 if edge $(i,j)$ is used in tour $k$}, \\
&v_{ik} \in \{0,1\} && \text{1 if vertex $i$ is visited in tour $k$}, \\
&u_i^k \in \{2,\cdots, n\} && \text{Visit order of vertex $i$ in tour $k$}, \\
&o_{ikk'} \in \{0,1\} && \text{1 if node $i$ is visited in both tours $k$ and $k'$},
\end{alignat*}

\noindent we then have the following integer linear problem:

{\small
% \begin{subequations}
\begin{align}
\max &\sum_{k=1}^{K} \sum_{i=2}^{n-1} \phi_i v_{ik} \tag{1} \\
\sum_{i=2}^{n} \psi_{i1}^k &= \sum_{j=2}^{n} \psi_{1j}^k \leq 1 &&\forall k \tag{2} \\
v_{jk} &= \sum_{i=1}^{n} \psi_{ij}^k = \sum_{l=1}^{n} \psi_{jl}^k &&\forall j = 2,\dots,n{-}1,\ \forall k \tag{3} \\
\sum_{i=1}^{n{-}1} \sum_{j=2}^{n} D_{ij} \psi_{ij}^k &+ \sum_{i=2}^{n} D_{i1} \psi_{i1}^k \leq D_{\max} &&\forall k \tag{4} \\
2 \leq u_i^k &\leq n &&\forall i=2,\dots,n,\ \forall k \tag{5} \\
u_i^k - u_j^k + 1 &\leq (n{-}1)(1{-} \psi_{ij}^k) &&\forall i \neq j,\ \ \forall k \tag{6} \\
o_{ikk'} &\leq v_{ik},\quad o_{ikk'} \leq v_{ik'} && \forall k', k < k' \tag{7} \\
o_{ikk'} &\geq v_{ik} + v_{ik'} - 1 && \forall k', k < k' \tag{8} \\
\sum_{i=2}^{n-1} o_{ikk'} &\leq T &&\forall k', k < k' \tag{9} \\
\psi_{ij}^k, v_{ik}, o_{ikk'} &\in \{0,1\},\quad u_i^k \in \{2,\cdots, n\} \tag{10}
\end{align}
% \end{subequations}
}

\subsection{An algorithm to solve the optimization model}
{\color{black}
The proposed framework uses a decomposition of the original problem. First, we consider a set of candidate tours obtained from the machine-learned risk map. Each tour is generated  to respect, by design, constraints $(2)$, $(3)$, $(4)$, $(5)$, $(6)$ and $(10)$, ensuring that it is a valid tour and that it respects the drone autonomy.

Let

\begin{equation}
\mathcal{C}=\{C_1,\ldots,C_m\}
\tag{11}
\end{equation}

be the set of candidate tours, and let $V(C_\ell)$ be the set of vertices visited by tour $C_\ell$. Once this set of candidate tours is obtained, we construct a conflict graph

\begin{equation}
G_c=(V_c,E_c). \tag{12}
\end{equation}

where

\begin{equation}
V_c=\mathcal{C}. \tag{13}
\end{equation}

Two tours are put in conflict when they overlap more than the threshold $T$:

\begin{equation}
(C_a,C_b)\in E_c
\Longleftrightarrow
|V(C_a)\cap V(C_b)|>T. \tag{14}
\end{equation}

Finding an independent set on this graph ensures that the overlap constraints $(7)$, $(8)$, $(9)$, $(10)$ are respected. Indeed, if $I\subseteq V_c$ is an independent set, then

\begin{equation}
(C_a,C_b)\notin E_c,
\qquad \forall C_a,C_b\in I,\ C_a\neq C_b. \tag{15}
\end{equation}

which is equivalent to

\begin{equation}
|V(C_a)\cap V(C_b)|\leq T,
\qquad \forall C_a,C_b\in I,\ C_a\neq C_b. \tag{16}
\end{equation}

Thus, finding an independent set of cardinality $K$ is equivalent to selecting $K$ mutually compatible tours from the candidate pool. This method produces feasible solutions by design. If the independent set returned by the solver has cardinality larger than $K$, we select the best subset of $K$ tours. If the returned independent set has cardinality smaller than $K$, the solution remains feasible. However missing an element of the optimal set of tours necessarily leads to suboptimal solutions. This selection problem is therefore a crucial step in the optimization process. We focus on the unweighted independent-set reduction because it captures the feasibility structure induced by the overlap constraints, while weighted physical encodings require additional normalization and local calibration choices that are implementation-dependent \cite{deOliveiraWMIS}. Nevertheless, incorporating weights directly in the independent-set solver is a promising direction for future work, since the current approach optimizes the expected reward only after feasible independent sets have been sampled, whereas a weighted encoding could bias the sampling process itself toward high-reward compatible tours.
}

We propose a specialized heuristic leveraging quantum computing protocols as a subroutine, named QUARTO (Quantum-Assisted Risk-based Tour Optimization). QUARTO proceeds through five steps: 1) Spatial Preprocessing, 2) Candidate Tour Generation, 3) Conflict Graph Construction, 4) Quantum-Assisted Optimization, and 5) Solution Reconstruction. \textcolor{black}{This method first requires the generation of high-quality candidate tours and then selects a spatially diverse subset of them. This could be achieved with multiple methods, and in this work we focus on the use of a quantum computer to tackle this subproblem. We demonstrate in this paper that a quantum routine can be integrated instead of classical routines, provided that the size of the conflict graph can be processed by the quantum device or emulator. A user solving the drone surveillance problem may then choose between QARTO or a classical version where step 4)'s subroutine is replaced by a classical solver.}

\begin{algorithm}[H]
\caption{QUARTO: Quantum-Assisted Risk-based Tour Optimization}
\label{alg:quarto}
\begin{algorithmic}[1]
\Require Probability map (grid or polygons); airbase coordinates; $K$ (desired tours); $S$ (sampled points per tour); $P$ (candidate tours pool size); $T$ (overlap threshold).
\Ensure Set of $k$ optimized UAV tours maximizing probability coverage.

\State \textbf{1. Spatial Preprocessing}
\Statex Convert the probability map into a graph $G=(V,E,w_V,w_E)$ where each vertex represents a cell/polygon center and each edge connects adjacent regions. Assign vertex weights $w_V$ as observation probabilities and edge weights $w_E$ as Euclidean or cost-based distances.

\State \textbf{2. Candidate Tour Generation}
\For{$\ell \gets 1$ \textbf{to} $P$}
  \State Select $S$ random (or high-probability) vertices from $G$ and create an auxiliary graph $G_\ell=(V_\ell,E_\ell)$.
  \ForAll{$(i,j)\in E_\ell$}
    \State Compute the shortest path on $G$ between candidates $i$ and $j$ using edge weights from the inverted probability map; denote its length by $d^{\mathrm{inv}}_{ij}$.
    \State Assign to edge $(i,j)$ in $G_\ell$ the tour-join cost
      $a_{ij} \gets d_{\text{airbase},i} + d^{\mathrm{inv}}_{ij} + d_{j,\text{airbase}}$,
      where $d_{\text{airbase},i}$ and $d_{j,\text{airbase}}$ are shortest-path distances to/from the airbase in $G$.
  \EndFor
  \State Find an approximate TSP tour on $G_\ell$ (using $a_{ij}$) to obtain candidate tour $\mathcal{T}_\ell$.
  \State Append $\mathcal{T}_\ell$ to the candidate list $\mathcal{C}_{\text{list}}$. \Comment Each edge $(i,j)$ in $G_\ell$ corresponds to the path found on $G$.
\EndFor

\State \textbf{3. Conflict Graph Construction}
\State Build $G_c=(V_c,E_c)$ where each vertex in $V_c$ represents a candidate tour from $\mathcal{C}_{\text{list}}$.
\State For any two tours, add an edge in $G_c$ if they share more than $T$ common waypoints.

\State \textbf{4. Quantum Heuristic (Maximum Independent Set)}
\State Map each vertex of $G_c$ to a qubit and embed the graph structure on the target device.
\State Apply the analog-quantum sequence implementing a MIS objective and sample bitstrings from the resulting quantum state.
\State Decode samples to candidate independent sets and select the best MIS solution $\text{MIS}^\star$ (or top few by value).

\State \textbf{5. Selection and Reconstruction}
\If{$|\text{MIS}^\star| \ge K$}
  \State Choose the subset of $K$ tours within $\text{MIS}^\star$ that maximizes total coverage value (forming a $K$-IS).
\Else
  \State Use $\text{MIS}^\star$ and, if needed, greedily add non-conflicting tours aiming to reach $K$, maximizing total value.
\EndIf
\State Construct the resulting set $\mathcal{C}_{\text{best}}$ from the selected tours; rebuild paths on $G$ for visualization/execution.
\State \Return Optimized set of $k$ tours with $ k \leq K$.
\end{algorithmic}
\end{algorithm}

\begin{figure}[H] 
    \centering
    \includegraphics[width=\columnwidth]{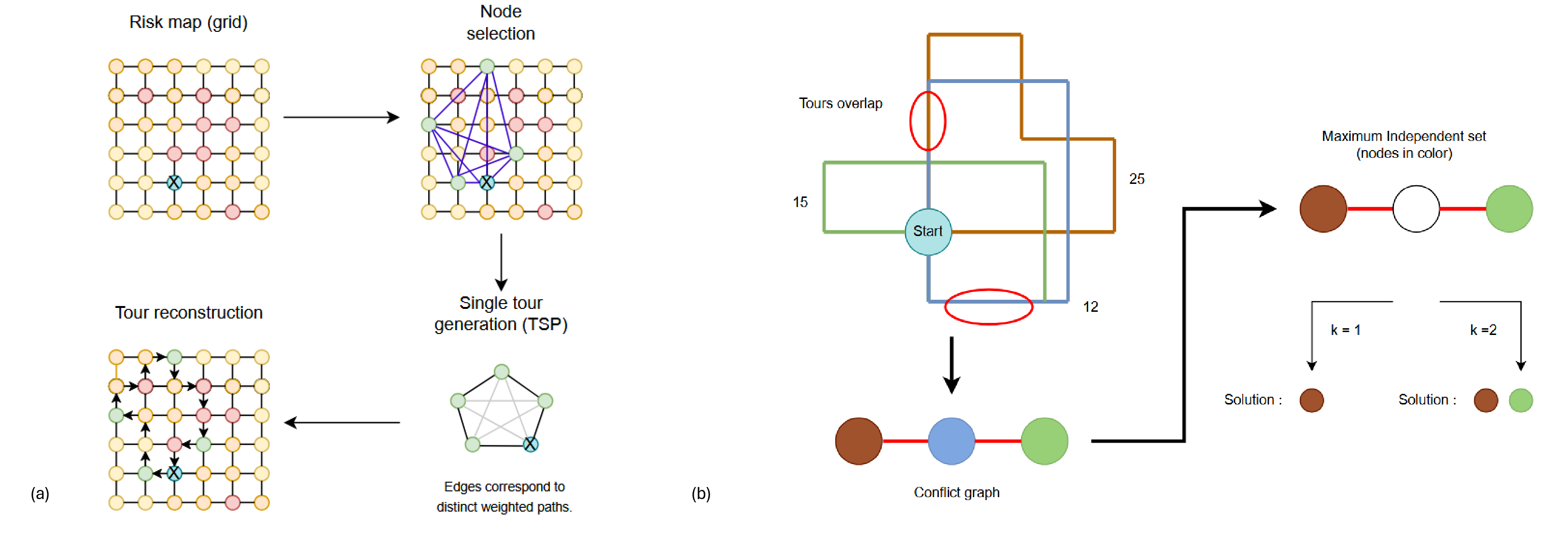} 
    \caption{(a) Illustration of the method to generate candidate tours. We first attributes to each node a weight $\phi_i$ which is the predicted ignition probability. A new edge weight that is the inverse of the sum of the ignition probability of the two vertices composing the edge is constructed $\tilde{d}_{ij} = (\phi_i + \phi_j)^{-1}$. We then select randoms nodes in the grid; each pair of those nodes corresponds to a shortest path using weight $\tilde{d}_{ij}$ as a metric. We then obtain a candiate tour using a randomized heuristic approach to the TSP on the resulting complete graph. The final tour is reconstructed from the individual paths - such a construction favours tours where the sum of total observed ignition probability is high. (b) An illustration of how the conflict graph is produced. Note that the conflict graph could be used for different values of $K$; here for example if a single tour is wanted ($K=1$) then the MIS could be used to select the brown tour, while if $K=2$, the same MIS could be used to return both nodes.}
    \label{fig:tour-construction}
\end{figure}

After the forecast map is converted into a graph $G$, we need to generate candidates tours on it. To do so an intermediate complete graph $G_I$ is defined, whose edge weights correspond to inverted probabilities from the risk map $\tilde{d}_{ij} = (\phi_i + \phi_j)^{-1}$, is built. Approximate solutions to the Traveling Salesman Problem (TSP) on random subgraphs of $G_l$ yields routes covering high-probability areas. As we do not seek the optimal solution of the TSP, we resort to the polynomial-time heuristic Christofides algorithm \cite{christofides2022worst}, which we randomize to provide a variety of tours. This mapping, from the original grid of estimated ignition probabilities to the candidate tours, is  visualized in Fig.~\ref{fig:tour-construction} (a).
The resulting tours may overlap; thus, a conflict graph is constructed $G_c$, where each node is a tour and an edge is present between tours if they overlap. This is visualized in Fig.~\ref{fig:tour-construction} (b).

A quantum heuristic then identifies a solution to the MIS on this graph. We use this MIS solution to extract a $K$-Independent Set ($K$-IS) if the size of the MIS is greater than $K$, we choose the $K$-Independent Set of maximum value among the resulting possibilities. In the case where $k$ is equal to the cardinal of the MIS, or less than this number, we use the MIS as it is. The resulting sets of vertices represents a set of $K$ non-conflicting tours (or less if $|\mathrm{MIS}| \leq K$), thus providing the final solution to the optimization sub-problem (see Sec.~\ref{sec:qopt} for details).

{\color{black}

\subsection{Complexity and optimality gap analysis}
\label{sec:complexity_gap_analysis}

The main combinatorial difficulty of the proposed pipeline appears in the final selection step. Once the candidate tours have been generated, selecting the best compatible subset is a maximum-weight bounded-$K$ independent set problem on the conflict graph. Because the set is bounded by $K$, a solution to this problem is an independent subset of at most size $K$ vertices. The vertices weights do not make the problem easier from a computational-complexity perspective: the unweighted case is recovered by setting all weights equal to one. Thus, the weighted case is at least as hard as the unweighted case. This unweighted selection problem is computationally hard because the classical independent set decision problem, which asks whether a graph contains an independent set of cardinality at least $K$, is NP-complete when $K$ is part of the input \cite{garey1979computers, karp1972reducibility}. In our formulation, $K$ is an arbitrary input parameter corresponding to the maximum number of tours allowed in the surveillance plan. Therefore, the induced unweighted $K$-IS subroutine remains computationally hard in general.

Let \(L\) denote the average number of visited vertices in a candidate tour. All transformations performed before the selection step are polynomial in the size of the discretized instance and in the number \(m\) of candidate tours. In particular, constructing the conflict graph requires comparing pairs of tours. If tours are stored as sets of visited vertices, this gives a cost of order \(O(m^2 L)\). These transformations therefore do not fundamentally change the complexity of the optimization process. They rather move the difficulty from the original routing formulation to a smaller, but still hard, selection problem.

The loss of optimality can be decomposed into two distinct terms. Let \(\mathrm{OPT}_{\mathrm{TOP}}\) be the optimal value of the full discretized TOP variant, let \(\mathrm{OPT}_{\mathcal{C}}\) be the optimal value restricted to the candidate pool \(\mathcal{C}\), and let \(\widehat{F}\) be the value returned by the implemented selection routine. Since the restricted problem optimizes over a subset of the feasible solutions of the full problem, we have
\[
\mathrm{OPT}_{\mathcal{C}} \leq \mathrm{OPT}_{\mathrm{TOP}}. \tag{17}
\]
The total optimality gap can then be written as
\[
\mathrm{OPT}_{\mathrm{TOP}} - \widehat{F}
=
\left(\mathrm{OPT}_{\mathrm{TOP}} - \mathrm{OPT}_{\mathcal{C}}\right)
+
\left(\mathrm{OPT}_{\mathcal{C}} - \widehat{F}\right). \tag{18}
\]
The first term is the candidate-pool gap. It measures the loss caused by restricting the search to the tours generated by the heuristic. The second term is the selection gap. It measures the loss caused by solving the induced $K$-independent set problem heuristically instead of exactly.

This decomposition also clarifies when the conflict-graph reduction is exact. If the candidate pool contains all tours of at least one globally optimal solution
\[
S^\star = \{C_1^\star,\ldots,C_r^\star\}, \qquad r \leq K, \tag{19}
\]
then \(S^\star \subseteq \mathcal{C}\) implies
\[
\mathrm{OPT}_{\mathcal{C}} = \mathrm{OPT}_{\mathrm{TOP}}. \tag{20}
\]
In that case, the only remaining gap comes from the independent-set selection routine.

Under a full-support assumption on the candidate tour generator, if each optimal tour \(C_i^\star\) is generated with probability \(p_i>0\) at each independent generation attempt. Let \(\mathcal{C}_N\) denote the set of candidate tours obtained after \(N\) generation attempts.
Then
\[
\mathbb{P}\left(C_i^\star \in \mathcal{C}_N\right)
=
1-(1-p_i)^N. \tag{21}
\]
Using the union bound, we obtain
\[
\mathbb{P}\left(S^\star \subseteq \mathcal{C}_N\right)
\geq
1-\sum_{i=1}^{r}(1-p_i)^N. \tag{22}
\]
Therefore,
\[
\mathbb{P}\left(\mathrm{OPT}_{\mathcal{C}_N} = \mathrm{OPT}_{\mathrm{TOP}}\right)
\geq
1-\sum_{i=1}^{r}(1-p_i)^N. \tag{23}
\]
This probability of finding all optimal tours belonging to one optimal set of tours converges to \(1\) as \(N \to \infty\) whenever \(p_i>0\) for all \(i\). This statement is not a finite-time approximation guarantee. It rather separates the candidate-generation error from the optimization error induced by the $K$-IS heuristic. The practical consequence of this analysis is that improving the probability of obtaining an optimal set of tours requires either increasing the probability that each optimal tour of a given optimal set is generated, or enlarging the candidate-tour pool. In other words, the performance of the method depends critically on the quality of the tour-generation procedure or on the availability of a sufficiently large conflict graph, if the tours generated are numerous but not of high quality.

}
\section{Quantum optimization} \label{sec:qopt}

\subsection{Optimization problems on neutral atoms computers}

A Quantum Processing Unit (QPU) is the generic name for any system that uses quantum resources to compute and process information. One such system is that of trapped neutral atoms in vacuum address by laser sources. In neutral atom QPUs, qubit states are encoded in the electronic ground state $\ket{g}$ (used to encode the logical state $\ket{0}$) and the highly excited Rydberg state $\ket{r}$ encodes the logical state $\ket{1}$. Atoms can be placed in a arbitrary two-dimensional register $\mathcal{R} = \{ \vec{r}_u \forall u \in V\}$ for $|V|$ atoms. Once atoms are placed, they are subject to strong interactions which strongly prohibit nearby atoms to both be in their respective Rydberg states - this is the Rydberg blockade mechanism \cite{Henriet2020quantumcomputing}. The dynamics of such a system under the effect of a nearly resonant laser source is governed by the Hamiltonian given by

\begin{align}
H(t) = \Omega(t) \sum_{u=1}^{|V|} \hat{\sigma}_u^x 
- \Delta(t) \sum_{u=1}^{|V|} \hat{n}_u 
+ \sum_{u < v = 1}^{|V|} U_{uv} \hat{n}_u \hat{n}_v \tag{24} \label{eq:hamiltonian}
\end{align}

First, \( \Omega(t) \) denotes the time-dependent Rabi frequency amplitude that governs the rate of transitions between atomic states. The term \( \Delta(t) \) represents the laser detuning (the laser frequency's offset from resonance between the $|r\rangle$ and $|g\rangle$ levels). This terms either promotes ($\Delta >0$) or disfavors ($\Delta <0$) the excitation of atoms. Given a set $V$ of atoms in the system, \( U_{uv} \) characterizes the interaction strength between atoms \( u \) and \( v \). It is given by $U_{uv} = C_6/r_{uv}^6$, where $C_6$ is a constant and $r_{uv} = || \vec{r}_u - \vec{r}_v||$ is the Euclidean distance between two atoms. Thus, nearby atoms strongly interact, and this interaction dies off very quickly as the distance is increased. The operators \( \hat{\sigma}^x_u \) correspond to the Pauli-X operator and \( \hat{n}_u \) corresponds to \( \hat{n}_u = (\mathbb{I} + \hat{\sigma}^z_u)/2 \), which counts presence of an excited atom, with $\hat{\sigma}^z_u$ the Pauli-Z operator. 

When the laser drive energy $\Omega$ is less than $U_{uv}$ for some pairs of atoms, we say that these atoms are blockaded - they cannot both be excited to their Rydberg states. By drawing edges between such pairs and by noting our atoms as the nodes of a graph, we thus obtain a unit-disk graph structure that is reconfigurable as the register is changed \cite{dalyac2024graph, bermot2025rydberg}. In this setting, neutral atoms QPUs can naturally tackle the Maximum Independent Set (MIS) problem by promoting a large positive detuning. The MIS consists in finding a maximal size set of non-adjacent vertices in a graph. If the set of non-adjacent vertices is instead not maximal but has \( K \) vertices, it is a \( K \)- independent set (\( K \)-IS). Any subset of the MIS is also an independent set. When controlling such a neutral atom QPU, information is only inputed in the system through three terms: \( U_{uv} \) (obtained from the positions of atoms, i.e. the register) and \( \Delta(t) \) and \( \Omega(t) \) which respectively reflect the temporal evolution of the detuning and the Rabi amplitude. Finding the good positions of atoms to represent a given graph problem in the QPU is often called embedding or register design, while finding the good amplitude and detuning evolution is called pulse-shaping. Both are crucial steps to ensure the final state we obtain represents a solution that corresponds to our input graph problem.

The combined effect of the inter-atomic interactions and the laser's drive gives rise to the time-dependent Hamiltonian shown in Eq.~\ref{eq:hamiltonian}, and in order to do computations, NAQCs thus manipulate the parameters of this Hamiltonian. This mode of processing quantum information is called analog quantum computing (AQC). Our quantum heuristic to sample valid solutions to the conflict graph is based on analog protocols.  
In analog quantum computing, we seek to find a time-dependent profile of $\Omega(t)$ and $\Delta(t)$ which will evolve an initial state \( |\psi_0\rangle \) to \( |\psi(t=t_f)\rangle = U(t_f) |\psi_0\rangle \), where $U(t_f) = \mathcal{T} \exp \left[ -i \int_0^{t_f} H(t') dt'\right]$ is the time ordered unitary evolution under a time-dependent Hamiltonian. Our objective is such that measurements of \( |\psi(t=t_f)\rangle \) in the computational basis (thus obtaining bitstrings) is the answer to some computational problem. In order to do that, we follow a conventional recipe \cite{coelho2022efficient, perron2025leveraging} where we fix \( \Omega(t = 0) = \Omega(t=t_f) = 0 \). At intermediate times, $\Omega$ reaches a maximal value $\Omega_{\max}$ which leads to prominent quantum fluctuations and an exploration of possible states. At \( t=0 \), we set \( \delta(t=0) = \delta_i < 0 \), and the instantaneous ground state is \( |0\rangle^{\otimes N} \), which is easily prepared. 
We prepare a given register \( \{ \vec{r}_u\} \) so as to represent the desired graph (an edge exists if $r_{uv} < R_b$, with $R_b = (C_6/\Omega_{\max})^{1/6}$) and choose a detuning \( \delta(t=t_f) =\delta_f > 0 \) accordingly, which will result in the final Hamiltonian having the MIS as its ground state. Our next task consists in proposing a pulse schedule \( \Omega(t) \) and \( \delta(t) \) that a) follows these constraints and b) prepares the instantaneous ground state of \( H(t_f) \) with the highest fidelity. Pulse shapes are generally inspired by the adiabatic theorem \cite{kato1950adiabatic, albash2018adiabatic}, which states that if the transformation of \( H(t) \) is slow enough, the system will always remain in its instantaneous ground state. In practice, this requires very long evolution times \( t_f \), well beyond the coherence time of the qubits in the system. Thus, a modified protocol called quantum annealing is used, where \( t_f \) is fixed but the pulse shape is distorted so as to minimize transitions from the instantaneous ground state to excited states, which would lower the final fidelity. Because of the loss of fidelity that such finite time analog protocol nevertheless incurs, multiple shots are generally needed to sample the final measured bitstrings, keeping only the one (or ones) that best help towards solving the combinatorial problem at hand.

In this section we provide details about the quantum heuristic that assists the whole framework. The motivation behind the use of this heuristic lies in the observation that we can take advantage of the constraint (9) which sets a maximum path similarity, to construct a conflict graph. A conflict graph is a graph where a vertex represents a possible drone tour and an edge represents an incompatibility between two tours. An incompatibility occurs in our problem when two paths are too similar and share more visited cells than the threshold $T$. Finding an independent set of size at most $K$, with $K$ the maximal number of paths allowed in one day (per operational constraints), guarantees to find a solution that respects the constraint (9). If the paths proposed in the conflict graph are of good quality, our method provides a set of at most $K$ paths as a valid high-value solution to the problem. 

We use the Python open access library \textsc{maximum-independent-set} \cite{pasqal2023mis, cazals2025quantum} (version 0.3.1), developed by Pasqal, to obtain a tentative MIS solution for a given graph $G$ using register and pulse design routines that design the analog quantum algorithm. Quantum protocols are emulated classically using the \textsc{emu-sv} open source library by Pasqal \cite{bidzhiev2025efficient}. We use the \textsc{maximum-independent-set} library like a black-box quantum solver for the combinatorial optimization problem and do not tune its configuration file. We use the obtained MIS solution as an input for the tour reconstruction step in our $K$-IS heuristic. The performance of the heuristic is dependant on the solution quality returned by the library, in a real world set-up it would also be dependant on the ability of the QPU to return high-quality solutions. The applications of ready-to-use quantum libraries is likely to develop as they provide an interface between models of problems and a quantum resolution method.

\begin{figure}[h]  % Forces the figure to appear "Here"
  \centering
  \includegraphics[width=0.5\columnwidth]{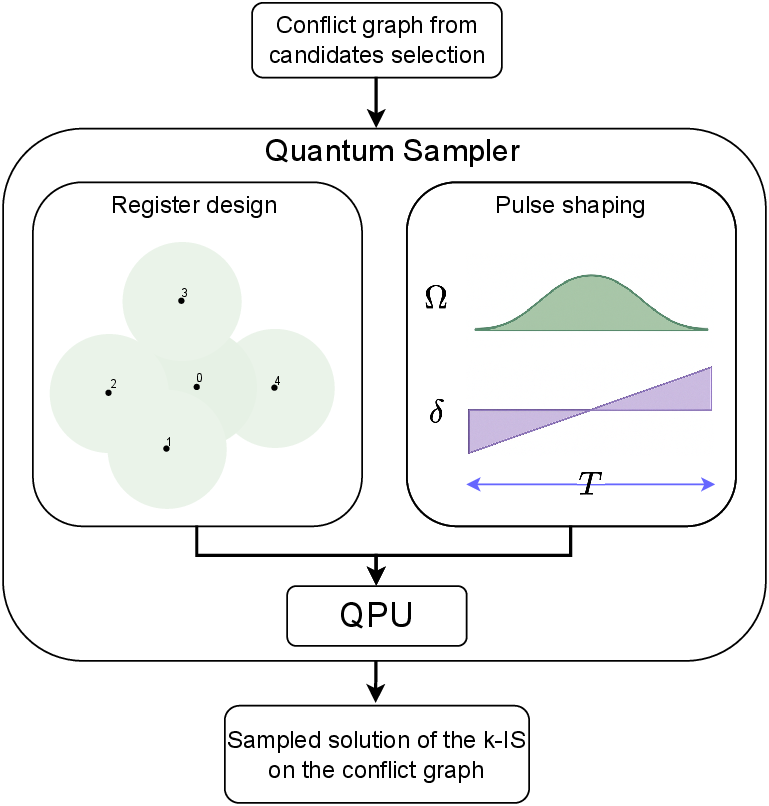}
  \caption{Description of the quantum heuristic. Additional steps of register design and pulse shaping are added; these respectively seek to find a faithful representation of the classical graph data in the quantum system and a good quantum protocol to obtain its solution. We create a conflict graph with incompatible candidates tours, then postprocess the resulting independent sets obtained with a quantum sampler to find a $K$-independent set proposed solution. 
  }
  \label{fig:my_image}
\end{figure}

\section{An application to wildfire prevention}\label{sec:wildfires}
As outlined above, we apply our framework to the use-case of wildfire prevention, where we are given historical forest fire geographical data on which we train the ML model, thus returning a graph optimization problem on which we apply a quantum heuristic. These steps are described in detail in the following subsections.  
These results demonstrate the effectiveness of our framework in returning feasible, actionable decisions quickly, whereas the optimal solution of the optimization problem cannot be found in the same time span.

\subsection{Data Selection and Study Area}

For this study, we utilize daily data spanning an eleven-year period (2010–2021) for the province of Québec, Canada. The geographical boundaries are defined by latitudes $45^\circ 15'00.0 \text{N}$ to $52^\circ 00'00.0 \text{N}$ and longitudes $79^\circ 30'00.0 \text{W}$ to $64^\circ 00'00.0 \text{W}$. Data points located in the ocean or outside Québec, including those in the United States or other Canadian provinces, are excluded, as drones must stay within a operational region. The size of this region was chosen to balance data volume and relevance, as it encompasses over $91\%$ of wildfires recorded in Québec during this period.
We focus on the period from March 11th to December 7th, as no wildfires were reported outside this timeframe across all 11 years. This selection ensures the consistency and reliability of our analysis by excluding outlier events.

\subsection{Datasets}
% \subsubsection{Wildfire and Weather Data}

Once the study area has been selected, we extracted our data from two datasets: wildfire records from the Canadian National Fire Database (CNFDB) \cite{CNFDB} and weather reanalysis data from the ERA5 product developed by the European Centre for Medium-Range Weather Forecasts (ECMWF) \cite{Hersbach2020ERA5}. The CNFDB provides historical data on fire occurrences, locations, and burned areas. Both datasets cover a the time frame 2010–2021.

ERA5 offers a detailed reconstruction of past atmospheric, land surface, and oceanic conditions, covering 1950 to the present, though we limit it to 2010–2021. It provides hourly data at a 31 km grid resolution. We extract ten key features: nine around midday—2-meter dewpoint temperature, 2-meter temperature, soil temperature at level 1, 10-meter wind components (u and v), surface pressure, leaf area index for high and low vegetation, and vegetation cover—plus total daily precipitation. All features are sampled at a $0.25^\circ$ resolution in both latitude and longitude.

\subsection{Data Processing}

Once these datasets are obtained, we process them into a usable format for our heuristic framework (see Alg.~\ref{alg:quarto}).

\subsubsection{Grid Construction}

We construct a grid over the study region, with each cell measuring $0.25^\circ$ in both latitude and longitude, as shown in Fig.\ref{fig:map}. This ensures alignment with the ERA5 dataset, where each grid cell corresponds to a central point.

\subsubsection{Wildfire Classification and Aggregation}

A cell is classified as "on fire" on a given day if at least one wildfire occurs within it; otherwise, it is labelled "not on fire". Wildfire data is aggregated daily within each cell, summing burned areas if multiple fires occur on the same day.

\subsubsection{Handling Excess Burned Area}

In rare cases (nine instances), the total burned area in a cell exceeds its actual size. When this occurs, the excess is redistributed to the nearest cells based on the Chebyshev distance:
\begin{align} 
\text{d}((x_1,y_1),(x_2,y_2)) = \max(|x_1-y_1|,|x_2-y_2|)  \tag{25}
\end{align} 
This adjustment has minimal impact, as in six cases, the excess area is transferred to a single neighbouring cell.

\subsubsection{Forecasting Labels}

For forecasting, we introduce an additional column where a value of 1 indicates that a fire will occur in the corresponding cell the following day, and a value of 0 indicates a prediction that no fire will occur. The dataset contains a total of $4,560,732$ rows, of which only $5,486$ are labelled with a 1. This represents approximately $0.0012\%$ of the dataset, highlighting a significant class imbalance.

\subsubsection{Partitioning}

The dataset is split such that data from 2010 to 2020 is used for training the model, while data from 2021 is reserved for testing. The data from 2021 is also used to evaluate the results of the quantum heuristic.

\subsection{Machine learning model}
Among all tested models, including logistic regression, decision trees, and random forests, Extreme Gradient Boosting (XGBoost) \cite{10.1145/2939672.2939785} consistently achieved the highest predictive accuracy. In practice, we found that its iterative refinement of decision trees and optimized feature weighting allows it to capture subtle patterns in wildfire occurrence. XGBoost also demonstrated strong generalization, reducing false positives while maintaining high recall in fire-prone areas. These observations confirm that ensemble methods like XGBoost are well-suited for wildfire forecasting, providing reliable probabilistic predictions to support prevention and response strategies.

Given the dataset’s imbalance, we trained the model on a balanced subset: all $4,911$ instances where a wildfire started the next day between 2010 and 2020, and $4,911$ randomly selected instances without a fire from the same period. This prevented the model from learning a biased "not on fire" majority class.

We then applied a 0.5 threshold to classify outputs as 'on fire' (1) or 'not on fire' (0), and evaluated performance against ground truth using standard metrics (see Table \ref{results}). To ensure a fair evaluation, the test set from 2021 included all 575 wildfire instances and 575 randomly sampled instances without a fire. On this balanced set, the model achieved a precision of $0.77$, recall of $0.83$, and F1-score of $0.79$ for wildfires. For cases without a fire, precision was $0.82$, recall $0.75$, and F1-score $0.78$. Overall accuracy reached $0.79$, indicating strong performance in detecting both classes under controlled conditions.

However, on the full 2021 dataset, which includes 379,137 non-fire cases, accuracy dropped to $0.73$. While precision for non-fire predictions approached 1, precision for wildfires fell to 0 due to a high number of false positives. Still, recall for wildfires remained high at $0.83$, meaning $83 \%$ of actual wildfires were detected.

This highlights the model’s strength in identifying rare yet critical events. Despite reduced precision in imbalanced settings, its high recall makes it a valuable tool for safety-first strategies. It reliably identifies wildfire risks, providing effective input for optimization algorithms that allocate limited prevention and mitigation resources.

\begin{table}[h]
\centering
\caption{The classification metrics (precision, recall, and F1-score) on a balanced test set $\mathcal{S}$ from the year 2021 ($|\mathcal{S}| = 575$). A high recall of 0.83 for class 1 means that most wildfires are correctly predicted.}
\label{results}
\begin{tabular}{|c|c|c|c|}
\hline
\textbf{Class} & \textit{Precision} & \textit{Recall} & \textit{F${}_1$-score} \\
\hline
0 & 0.82 & 0.75 & 0.78 \\
\hline
1 & 0.77 & 0.83 & 0.79 \\
\hline
\end{tabular}
\end{table}

\subsection{Quantum optimization}

\subsubsection{\textcolor{black}{Implementation}}

\begin{figure}[!t]
\centerline{\includegraphics[width=\columnwidth]{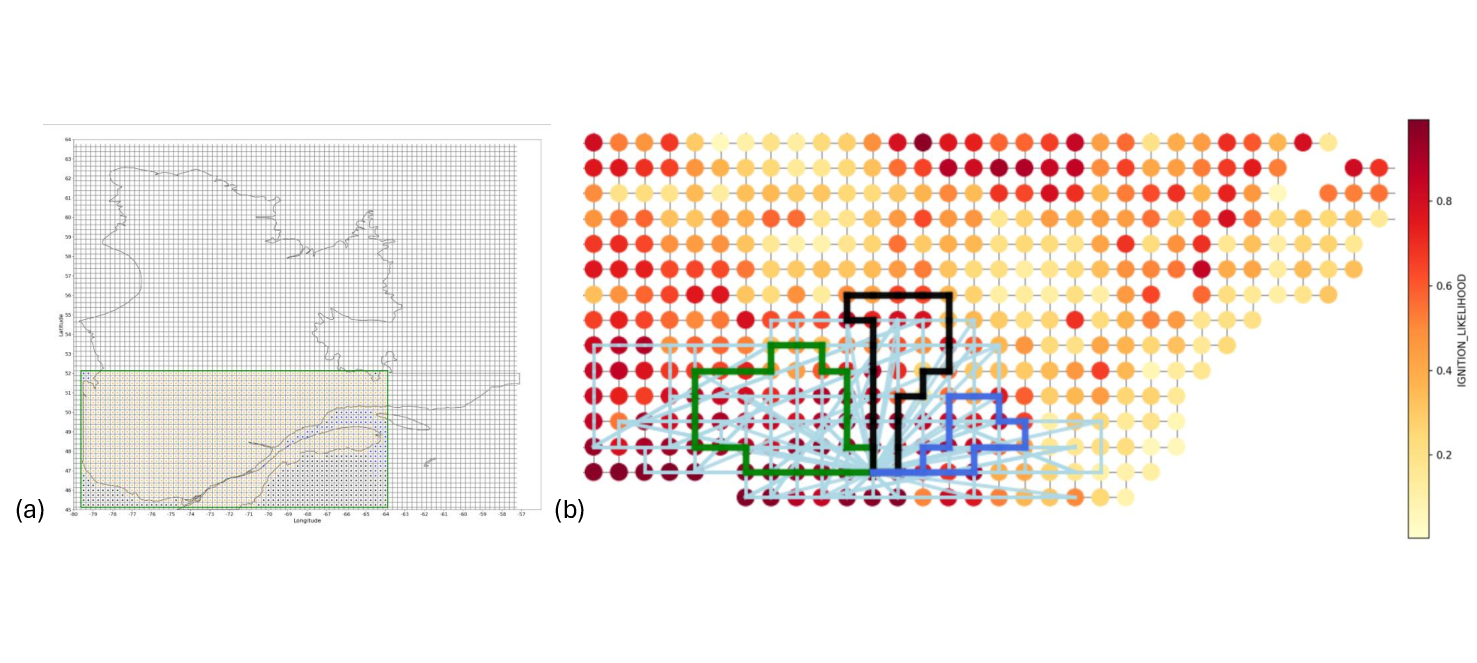}}
\caption{(a) The map of Québec is divided into grid cells, each measuring $0.25^\circ$ in both latitude and longitude. Over $91\%$ of wildfires occur within the area outlined in green. However, our dataset only considers cells containing an orange dot, as the blue and black cells are either located in the ocean or in regions outside Québec. (b) A solution for 3 tours (maximum length of 500km) within a day (June 1st, 2021) and a threshold of at most 3 commons points between tours in the conflict graph construction. In light blue is the set of candidates tours, while in black, green and blue is the set of selected tours. We can see that both spatial diversity and likelihood observation are respected. }
\label{fig:map}
\end{figure}

The grid map of ignition probabilities $\phi_i$ is used to propose tours and thus define the conflict graph on which a $k$-IS is used for the heuristic solution to the drone schedule (see Algo.~\ref{alg:quarto}). Our quantum heuristic is programmed using the library \textsc{maximum-independent-set} \cite{pasqal2023mis, cazals2025quantum} provided under a MIT-like license by the company Pasqal. We set the maximal size of our conflict graphs at 10, 15 and 20 (corresponding to the number of qubits we will use) in order to test the change in performance of our heuristic when given access to a larger number of qubits. We set the number of drone tours per day to 5, a number that could be reached with 2 to 3 drones, which is a realistic operational scenario given the current price of long-range drone. The similarity threshold between two tours is set to 30 \% - if their paths overlap by more than that proportion, an edge in the conflict graph is present. We set the autonomy for 1 tour to 500 km. Fig.~\ref{fig:map}  shows an example visualization of the different candidate tours (light blue) with the three drone tours selected by the quantum heuristic displayed in bold. The heat map corresponds to the ignition probability map obtained from the ML model for that day. 

As we do not yet have access to quantum hardware to test our framework, we deployed it on classical emulators that seek to replicated the full analog quantum dynamics and allow us to integrate them with the otherwise classical workflow. The numerical tests are done using the EMU-SV emulator \cite{bidzhiev2025efficient}, run on a 2 x AMD EPYC 7643 milan (total 96 cores) architecture with 500 Gbs of memory. When emulating the quantum protocol, we ask for 100 bitstrings to be returned, corresponding to the repetition and measurement of the quantum protocol of 100 times. Given the current performance of NAQC QPUs, one can expect an effective shot rate of $\sim 1$ Hz. In practice, this means that the quantum part of our heuristic can be done in approximately one minute and thirty seconds on a real QPU. Note that current roadmaps envision that this shot rate can go as high as $1000$Hz with future hardware developments~\cite{dalyac2024graph}.

\subsubsection{\textcolor{black}{Results and comparison to a generalist solver}}

\sout{Although} Our method is highly flexible and will work even for very small quantum resources (approximately $\frac{3}{2}k$ physical qubits, with $k$ the number of drone tours). The expected performance improves with larger conflict graphs and thus with the number of associated qubits. Small conflict graphs imply a narrower exploration of the solution space of possible tours, which leads to graphs built with few quantum resources to produce worse solutions on average than graphs built with a larger number of quantum resources (i.e. more nodes). 

\begin{figure}[h]
\centerline{\includegraphics[width=\columnwidth]{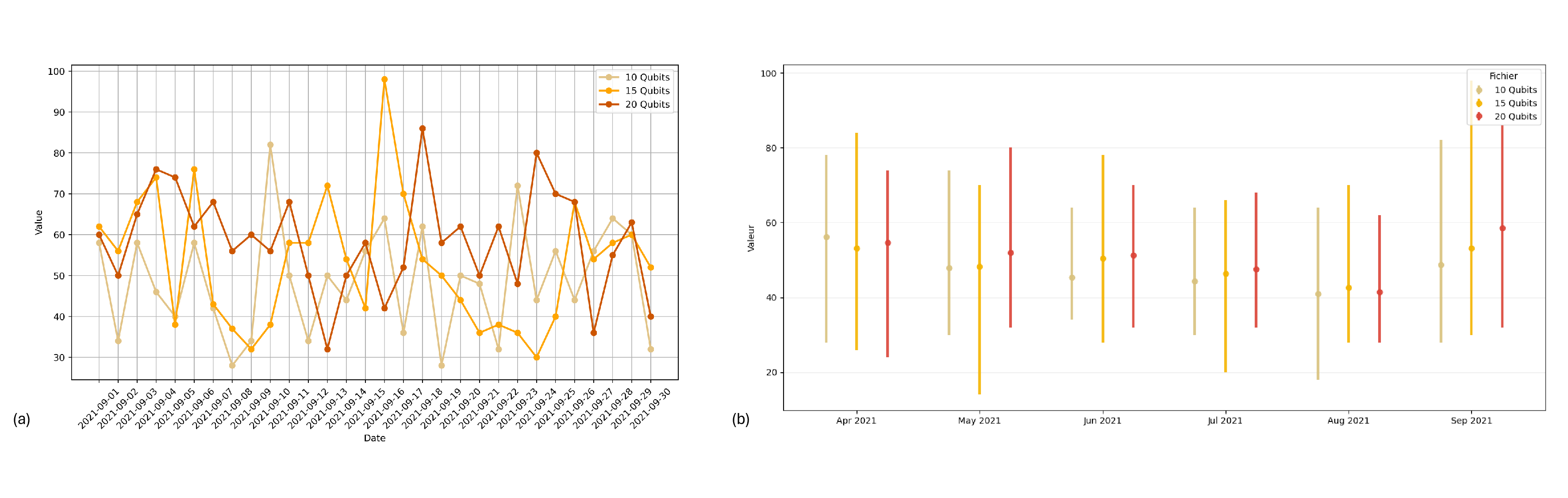}}
\caption{
(a) Comparison of the heuristic's objective value (sum of local ignition probabilities along the proposed drone tours) for September 2021 as a function of the quantum resources we allow to the conflict graphs (10, 15, 20 qubits). On average, more resources leads to better results. (b) Comparison of the average of the heuristic's objective value over a month, for the six month of 2021 with active fire risks. all months with a finite wildfire risk depending on the quantum resources we allow to the conflict graphs (10, 15, 20 qubits). The bars display the average (center point), minimum (lower bracket) and maximum (upper bracket) for each month. Though the quantum resources are limited, the heuristic's performance is improved on average with the higher number of qubits corresponding to more tours being considered at the same time and thus more exploration.
}
\label{results-sept}
\end{figure}

This is observed in practice in Fig.~\ref{results-sept} and in particular Fig.~\ref{results-sept} (b). In these figures, we show the objective value $\sum_k\sum_{i\in \mathcal{P}_k} \phi_i$ (i.e. the cumulative sum of local ignition probabilities) reached daily for the month of September 2021 (see Fig.~\ref{results-sept} (a) ) and the average objective value per month for the year 2021 (see Fig.~\ref{results-sept} (b) ). Fig.~\ref{results-sept}(a) shows the stochastic nature of heuristics when the number of qubits (i.e. number of nodes in the conflict graph) is increased. Although it seems at first glance that no curve dominates the other, the average value for 20 qubits curve (red) is 51.935, which is superior to the 15 qubits curve (orange; average of 48.194) and 10 qubits curve (yellow; average of 47.968). Adding more qubits thus leads on average to better results. This observation is confirmed in Fig.~\ref{results-sept} (b), where we display the monthly average for 6 consecutive months, minimum and maximum values for 10, 15 and 20 qubits. 

We finally compare the heuristic framework we propose to a state-of-the-art solver, namely the HiGHS \cite{huangfu2018parallelizing} classical solver. It is an open-source optimization solver that provides state-of-the-art performance for large-scale linear programming (LP), mixed-integer programming (MIP), and quadratic programming (QP). This solver seeks to find the optimal solution to the MILP outlined in Sec.~\ref{optModel}.

\begin{table}[h]
\caption{Comparison of a classical solver and heuristic performance over July 2021, showing the solver's low reliability for a realistic time budget and the heuristic’s consistent valid outputs.}
\label{tab:solver-heuristic}
\centering
\begin{tabular}{lcc}
\toprule
\textbf{Metric}            & \textbf{Solver (HiGHS)} & \textbf{Algo. 1} \\
\midrule
Valid outputs              & 10      & 31    \\
Failure rate               & 63\%    & 0\%   \\
Minimum value (valid only) & 0.08    & 32    \\
Average value (valid only) & 0.53    & 47.19 \\
Maximum value (valid only) & 3.84    & 68    \\
Solving time (s)           & 3600    & 120   \\
\bottomrule
\end{tabular}
\end{table}

To be realistic we allow it a 1 hour time budget, as the drones have to fly for the whole day and the route schedule has to be given quickly. We compare only solving time for both methods, the modelling time for the solver is not counted. For the quantum heuristic we consider the solving time to be the time spent using the QPU. We see that the solver is unable to return valid outputs a majority of the time, and that when it returns a valid output (i.e. a selection of tours), the quality of these is wildly lower than that of the quantum heuristic we propose in this paper.

\subsubsection{Evaluating against specialized K-IS heuristics}
\label{sec:selection-comparison}
{\color{black}
The comparison with a general-purpose solver remains relevant in a context where no specialization is used. Such a situation can occur when a quick implementation is needed, since practitioners may first model the problem directly and solve it with an off-the-shelf optimizer. However, this comparison is not sufficient to evaluate the selection component itself, since one may also implement well-known heuristics or a specialized formulation for the induced $K$-IS problem.

We therefore add an ablation study restricted to the optimization component of the framework. The candidate pool, tour weights, and conflict graph are held fixed. This allows us to evaluate how the quantum-assisted selection routine behaves inside the pipeline relative to classical selection methods. In particular, we compare the quantum-assisted routine used in QUARTO with random compatible selection, greedy $K$-independent-set heuristics \cite{SAKAI2003313}, simulated annealing \cite{kirkpatrick1983optimization}, brute-force enumeration when available, and a specialized MILP formulation solved with HiGHS \cite{huangfu2018parallelizing}.

For this comparison, the restricted maximum-weight $K$-independent-set problem is formulated with one binary variable $x_\ell$ for each candidate tour $T_\ell \in \mathcal{C}$. The MILP maximizes the selected tour weight, under the incompatibility constraints induced by the conflict graph and the cardinality constraint:
\[
\max \sum_{T_\ell \in \mathcal{C}} w_\ell x_\ell \tag{26}
\]
subject to
\[
x_a + x_b \leq 1, \qquad \forall (C_a,C_b) \in E_C, \tag{27}
\]
\[
\sum_{C_\ell \in \mathcal{C}} x_\ell \leq K, \tag{28}
\]
\[
x_\ell \in \{0,1\}, \qquad \forall C_\ell \in \mathcal{C}. \tag{28}
\]
The LP relaxation is obtained by replacing the integrality constraints with
\[
0 \leq x_\ell \leq 1, \qquad \forall C_\ell \in \mathcal{C}. \tag{29}
\]
Its optimal value $U_C^{LP}$ provides a valid upper bound on the restricted optimum $\mathrm{OPT}_{\mathcal{C}}$. When the binary formulation can be solved to proven optimality, the exact restricted optimum is also reported. For instances that cannot be solved exactly within the prescribed time limit, the LP-relaxation value provides a common reference upper bound for all selection methods. The reported metrics include the objective value, the LP-bound gap, the feasible rate (proportion of returned feasible tours) and the runtime. The LP-bound gap is defined as
\[
\delta_A^{LP}
=
\frac{U_C^{LP}-\hat{F}_A}{U_C^{LP}}, \tag{30}
\]
where $\hat{F}_A$ is the objective value returned by method $A$. 
In order to study precisely the selection step, we fix the tours on which the optimization is performed. All tours and grid maps are made available in the reproducibility package. We randomly choose to study the September 30, 2021 instance.

\begin{table}[h]
\centering
\color{black}
\setlength{\tabcolsep}{1pt}
\renewcommand{\arraystretch}{0.99}
\caption{Selection-stage comparison for the September 30, 2021 instance, using conflict graphs ($C_G$) of 10, 15, and 20 vertices.}
\label{tab:selection_art_all}
\begin{tabular}{llcccc}
\toprule
$C_G$ size & Selection method & Objective & LP-bound gap & Feasible rate & Runtime \\
\midrule
\multicolumn{6}{l}{\textbf{10}} \\
 & Random compatible selection & 19.26 & 56.76\% & 100\% & 0.000038 \\
 & Greedy weighted K-IS & 21.18 & 52.43\% & 100\% & 0.000048 \\
 & Greedy weight/degree K-IS & 21.18 & 52.43\% & 100\% & 0.000029 \\
 & Greedy with local improvement & 21.18 & 52.43\% & 100\% & 0.000084 \\
 & Simulated annealing & 21.18 & 52.43\% & 100\% & 0.044 \\
 & Brute-force exact & 21.18 & 52.43\% & 100\% & 0.002 \\
 & Exact/time-limited MILP & 21.18 & 52.43\% & 100\% & 0.009 \\
 & Quantum-assisted routine & 20.74 & 53.42\% & 100\% & 100$^{*}$ \\
\midrule
\multicolumn{6}{l}{\textbf{15}} \\
 & Random compatible selection & 21.33 & 56.09\% & 100\% & 0.000050 \\
 & Greedy weighted K-IS & 21.18 & 56.39\% & 100\% & 0.000061 \\
 & Greedy weight/degree K-IS & 21.18 & 56.39\% & 100\% & 0.000043 \\
 & Greedy with local improvement & 27.34 & 43.71\% & 100\% & 0.000201 \\
 & Simulated annealing & 21.18 & 56.39\% & 100\% & 0.077 \\
 & Brute-force exact & 28.37 & 41.61\% & 100\% & 0.021 \\
 & Exact/time-limited MILP & 28.37 & 41.61\% & 100\% & 0.010 \\
 & Quantum-assisted routine & 28.37 & 41.61\% & 100\% & 100$^{*}$ \\
\midrule
\multicolumn{6}{l}{\textbf{20}} \\
 & Random compatible selection & 21.06 & 56.73\% & 100\% & 0.000069 \\
 & Greedy weighted K-IS & 21.18 & 56.48\% & 100\% & 0.000087 \\
 & Greedy weight/degree K-IS & 27.14 & 44.24\% & 100\% & 0.000074 \\
 & Greedy with local improvement & 27.14 & 44.24\% & 100\% & 0.000170 \\
 & Simulated annealing & 27.14 & 44.24\% & 100\% & 0.136 \\
 & Brute-force exact & 27.14 & 44.24\% & 100\% & 0.146 \\
 & Exact/time-limited MILP & 27.14 & 44.24\% & 100\% & 0.011 \\
 & Quantum-assisted routine & 21.25 & 56.34\% & 100\% & 100$^{*}$ \\
\bottomrule
\end{tabular}

\vspace{0.5em}
\raggedright
\footnotesize{$^{*}$ The theoretical time on a NAQC is 100 seconds, corresponding to 100 measurements at 1 Hz. Emulation time is available in the reproducibility package.}
\end{table}

{\color{black}
The results show that all methods return feasible selections. For the conflict graph of size 10, the quantum-assisted routine is slightly below the best classical methods. For the conflict graph of size 15, it reaches the best value found by the exact and MILP methods. For the conflict graph of size 20, it does not recover the best classical solution and stays close to the simpler baselines. These results do not show a systematic advantage of the quantum-assisted routine. Rather, they show that it can be inserted as a feasible selection component in the pipeline, while its quality  remain vastly dependent on the instance and the resulting conflict graph. For runtime estimations, we fixed a setting where the quantum computer's measured output is obtained 100 times. Current NAQCs can achieve a shot rate of $\sim 1$ Hz, though future hardware improvements can increase this by up to 2 orders of magnitude \cite{dalyac2024graph}. Since this selection step is modular, any improvement of the independent-set routine, whether classical or quantum, would be directly transmitted to the full method, as long as the candidate pool and reconstruction steps remain unchanged. This limited exploration confirms that any benefit of a quantum computing heuristic in the pipeline would only arise in cases where the conflict graph is large, as was shown in Ref.~\cite{cazals2025quantum}.
}
}

\FloatBarrier
\section{Concluding remarks}\label{sec:conclusion}

This work motivates the development of new quantum heuristics for Orienteering Problems and opens avenues for integrating quantum algorithms into broader optimization pipelines. In this paper, we demonstrate that quantum optimization can be integrated into a useful decision-making framework, showcasing a real-world example in the context of wildfire management. This represents a step toward the practical deployment of quantum-assisted optimization pipelines in high-impact applications. In particular, \textcolor{black}{our results show that neutral-atom-compatible quantum routines can be used as heuristics for the $K$-IS selection problem}, which can itself be used as a subroutine in a larger heuristic framework for drone surveillance tasks, which we detail in this paper. At the scale of our considered problems (maximum 20 qubits), competing classical techniques are faster compared to currently available quantum computers. However, such quantum computers are rapidly scaling upwards, and access to hundreds to thousands of coherently controlled neutral atoms \cite{dalyac2024graph} would allow the consideration of a large number of candidate tours on a single QPU, \textcolor{black}{which can reduce the candidate-generation error identified in Section~\ref{sec:complexity_gap_analysis}}. For very large conflict graphs, the analog quantum process that lies at the heart of our heuristic leads to a distribution  \textcolor{black}{of bitstrings from which we extract a solution to the selection subproblem}. Emulating classically this distribution is very hard~\cite{bidzhiev2025efficient} for more than 50 qubits. This challenge makes it not so trivial to perform full benchmark comparisons using only classical quantum simulation tools. It is also in this regime that analog quantum computers are thought to outperform classical methods~\cite{Ebadi2022MIS}. Given this, we view our optimization method as a whole pipeline, the same way a solver is viewed. This led us to compare it directly with the solver HiGHS.  \textcolor{black}{If we consider only the selection step of the framework we can} perform internal benchmarking of the inner workings of the selection heuristic where we replace the quantum subroutine with alternatives.  \textcolor{black}{This shows that the quantum method does not degrade the feasibility of the pipeline and can obtain values comparable to some classical heuristics on selected instances.} A comparison with other quantum heuristics, such as QAOA~\cite{farhi2014quantumapproximateoptimizationalgorithm}, is left for future work. The framework in itself is a contribution to the field of wildfire management and, in general, to the task of optimizing aerial surveillance routes. 
Our framework illustrates the capability of neutral atoms quantum processors to tackle combinatorial problems in conjunction with machine learning. We used XGBoost as a machine learning model to predict the probability that a given region will see a forest fire start in the next day. We note that quantum-classical machine learning approaches, such as that of Ref.~\cite{leclerc2024financialriskmanagementneutral} could also be envisioned there. Our results suggest the potential of hybrid quantum algorithms — particularly heuristics — to address specific logistics challenges. Notably, we emphasize that we used only openly accessible software tools to specify the analog quantum process and emulate it. This should encourage others to consider these tools in the design of routing heuristics and the integration of quantum routines within them. 

\section*{Declarations}

\bmhead{Funding}

The authors acknowledge financial support from an Alliance grant (grant number ALLRP 590810 - 23) funded by the Natural Sciences and Engineering Research Council of Canada (NSERC) and Pasqal Canada. This work made use of compute resources by Calcul Québec and the Digital Research Alliance of Canada.

\bmhead{Data availability}
The ERA5 reanalysis dataset provided by ECMWF was used (time frame 2010–2021) and is publicly accessible via the Copernicus Climate Data Store (https://cds.climate.copernicus.eu). The Canadian National Fire Database (CNFDB, 2010–2021) was used and is accessible via Natural Resources Canada (https://cwfis.cfs.nrcan.gc.ca/ha/nfdb). The processed datasets generated in Sec.~\ref{sec:selection-comparison}, along with the analysis code, are available at: \url{https://github.com/QUASAR-UdS/TOP/}.

\bmhead{Conflict of interest}
DL, AV and YBL declare no conflict of interest. VDT was a part-time employee at Pasqal Canada during the initial writeup of the paper.

\bibliography{biblio}% common bib file

\end{document}